\documentclass{mn2e}
\usepackage{graphicx}

\title{Circumstellar interaction in type Ibn supernovae and SN~2006jc}
 
\author[N.N.~Chugai]
{N.N.~Chugai \thanks{nchugai@inasan.ru}\\ 
Institute of Astronomy, RAS, Pyatnitskaya 48, 119017 Moscow,
 Russia\\
}

\date{Accepted 2009
      Received 2009;
      in original form 2009}

\pagerange{\pageref{firstpage}--\pageref{lastpage}}
\pubyear{2009}

\begin{document}

\maketitle

\label{firstpage}

\begin{abstract}

I analyse peculiar properties of light curve and continua of 
enigmatic Ibn supernovae, including 
SN~2006jc, and argue in favour of the early strong circumstellar 
interaction. This interaction explains the high luminosity and fast flux  
rise of SN~1999cq, while the 
cool dense shell formed in shocked ejecta can explain the smooth early 
continuum of SN~2000er and unusual blue continuum of SN~2006jc. 
The dust is shown to condense in the cool dense shell at about day 50. 
Monte Carlo modelling of the He\,I 7065 \AA\ line profile affected 
by the dust occultation supports a picture, in which 
the dust resides in the fragmented cool dense shell, whereas  
He\,I lines originate from circumstellar clouds shocked and fragmented 
in the forward shock wave.

\end{abstract}


\section{Introduction}

Recently a new family of supernovae (SN), so called SN~Ibn, 
came in the limelight (Matheson et al. 2000; Pastorello et al. 2008). 
These objects, showing strong narrow (FWHM$\sim 2000$ km s$^{-1}$) He\,I 
emission lines and week or no H$\alpha$ emission, are suggested to 
be SN~Ibc interacting with a helium-rich circumstellar 
matter (CSM) (Foley et al. 2007; Pastorello et al. 2007; 
Pastorello et al. 2008). 
Four known members of this family are: SN~1999cq (Matheson et al. 2000), 
SN~2000er and SN~2002ao (Pastorello et al. 2008), and well
studied SN~2006jc (Nakano et al. 2006; Foley at al. 2007;
Pastorello et al. 2007; Di Carlo et al. 2008; 
Smith et al. 2008; Anupama et al. 2008). 
In SN~2006jc the CS interaction is manifested also 
by X-ray emission (Immler et al. 2008).  
At the position of SN~2006jc an optical transient 
was detected in Oct. 2004 with the maximal absolute magnitude 
 of $M_R\sim-14$ (Nakano et al. 2006; Foley et al. 2007;
 Pastorello et al. 2007). This outburst is 
 blaimed for the mass ejection by presupernova (Foley at al. 2007; 
Pastorello et al. 2007).

A striking feature of SN~2006jc is the early dust formation 
 ($t\sim 50$ d) indicated by the infrared 
emission of hot ($T\sim1800$ K) dust after about day 50 
(Di Carlo et al. 2008; Smith et al. 2008; 
Mattila et al. 2008) and 
 blue shift of emission lines (Smith et al. 2008; 
Mattila et al. 2008). Although 
dust might form in the unshocked ejecta as shown by  
Nozawa et al. (2008), the dust formation in the shocked cool 
dense shell (CDS) seems to be preferred (Mattila et al. 2008). In fact, 
dust condensation in the CDS  has been invoked already for 
type IIL supernova 1998S (Pozzo et al. 2004).

A general wisdom is that the light of SN 2006jc is powered by 
the radioactive decay of $^{56}$Ni similar to normal SNe~Ibc 
(Tominaga et al. 2008). The X-ray emission is 
explained by Tominaga et al. (2008) as a result of 
interaction of SN~Ibc ejecta with smoothly distributed 
 rarefied CSM with the total mass of $\sim0.003~M_{\odot}$. 
 In this model both reverse and forward shocks are adiabatic, 
 i.e., the dust formation in the shocked gas is precluded.
In scenario proposed by Mattila et al. (2008) 
 SN~2006jc ejecta collide with a massive $\sim0.5~M_{\odot}$ 
CS shell at the distance of $1.5\times10^{16}$ cm. 
The massive CDS formed in the forward shock was 
presumably the site where the dust condensed after day 50 
(Mattila et al. 2008). Unfortunately authors did not discuss
a possible contribution of CS interaction in the bolometric 
luminosity. 

The present study is motivated by the unsettled issue of the dust-forming 
site and striking facts presently escaped the detailed analysis, viz.,  
(a) luminous maximum and fast pre-maximum flux rise of SN~1999cq;
(b) smooth early continuum of SN~2000er; and (c) odd
 blue continuum of SN~2006jc.
The goal of the paper is to propose a model 
that could account for the above properties and some
other observational data, including the dust formation. 
The paper is organized as follows. Arguments in 
favour of a strong CS interaction in SNe~Ibn at the early stage 
are considered in Section \ref{sec-general}.
The interaction models with and without $^{56}$Ni are then applied 
to account for the bolometric light curve, X-ray emission, 
 and moderate expansion velocity of 
outer layers (Section \ref{sec-interact}). 
The parameters of the CDS recovered in 
these models are used to study the dust formation in terms of the 
theory of homogeneous nucleation (Section \ref{sec-dust}).
The Monte-Carlo simulations of 
dust occultation effects in He\,I lines are made to illustrate that 
the proposed general picture is sensible (Section \ref{sec-bshift}).
The results and some worrisome issues 
are discussed in Section \ref{sec-discuss}.

Hereafter we adopt for SN~2006jc the explosion date 2006 Sep. 21 or 
JD=2454000 (Mattila et al. 2008) and the distance of 26 Mpc 
according to redshift from the Lyon Extragalactic
 Database and assuming $H_0=70$ km s$^{-1}$ Mpc$^{-1}$.

\section{Early CS interaction}
\label{sec-general}

Pastorello et al. (2008) emphasise the remarkable homogeneity in the 
spectral properties among SNe~Ibn. This fact provides us a confidence 
that the conclusions made from the analysis of particular objects 
can be applied to other members of the family. 
Here I consider observational data which suggest a dominant 
contribution of CS interaction in the early luminosity 
and spectra of SNe~Ibn.

\subsection{Light curve and CSM}
\label{sec-gen1}

Unusually high luminosity of SN~1999cq (type Ibn) at the 
light maximum with the 
absolute magnitude $M_R\sim -19.9$ mag (Matheson et al. 2000; 
Pastorello et al. 2008) and the fast pre-maximum 
flux rise by $>3$ mag in four days (Matheson et al. 2000) 
are unprecedented for SNe~Ibc including luminous SN~1998bw.
Both high luminosity and fast brightening suggest that early 
SN~1999cq was powered 
by the ejecta interaction with a dense CSM, a scenario 
similar to that proposed for the early light curve of 
type IIL SN~1998S (Chugai 2001). SN~2006jc was detected on day 17 
after the adopted explosion date (Nakano et al. 2006; Pastorello et al. 2008) 
and its absolute magnitude at the discovery was similar to that of 
SN~1999cq at the same age. It seems plausible, therefore, that the 
prediscovery light curve of SN~2006jc was similar and the early 
luminosity of SN~2006jc was also affected by the CS interaction.
The early interaction should lead to the deceleration of outer 
layers of ejecta which is supported by the fact that 
the early ($t<10$ d) 
He\,I 5876 \AA\ emission of SN~2000er shows the maximal 
expansion velocity of only $\sim 9000$ km s$^{-1}$ 
(Pastorello et al. 2008), rather moderate value for early SN~Ibc.

The total energy radiated by SN~1999cq during the first 10 days 
after the light maximum is about $E_{r}\sim2\times10^{49}$ erg 
according to the available light curve  (Matheson et al. 2000; 
Pastorello et al. 2008).
To produce that energy a shock wave with velocity of 
$10^4$ km s$^{-1}$ had to interact with the CS mass of about 
$M_{cs}\sim0.02~M_{\odot}$ assuming 100\% radiation efficiency.
For that CSM mass enclosed 
within $r\sim vt\sim10^{15}$ cm the density parameter for the steady 
outflow turns out to be $w=M_{cs}/r\sim 4\times10^{16}$ g cm$^{-1}$. 
For the density distribution in outer SN ejecta $\rho\propto v^{-9}$ 
the shocked ejecta should be by 2.5 times more massive (Chevalier 1982b), 
so the expected CDS mass on day 10 could attain $\sim0.05~M_{\odot}$.

\begin{figure}
\includegraphics[width=80mm]{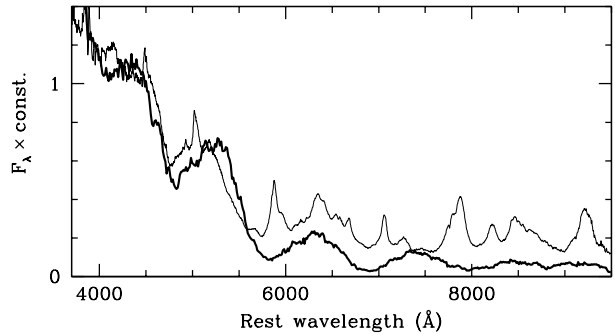}
\caption{
Blue continuum of SN~2006jc on day 22.
The model of line emission from cool dense shell (thick solid line) with the 
reddening $A(V)=0.15$ is overplotted on the observed 
spectrum (Fooley et al. 2008).
}
\label{f-synt}
\end{figure}

\subsection{Early smooth continuum of SN~2000er}
\label{sec-gen2}

The earliest spectrum of 
 SN 2000er shows a smooth continuum (Pastorello et al. 2008), 
very unlike bumpy continua of any early SN~Ibc. I suggest 
that the smooth continuum forms in a thin CDS 
likewise in early SN~1998S (Chugai 2001).
The smooth continuum of SN 2000er grows bumpy
on the time scale of five days after the first spectrum 
(Pastorello et al. 2008), which 
means that the continuum forms in a partially transparent regime, 
i.e., the optical depth is $\tau\leq1$. 
For the luminosity of $\sim10^{43}$ erg s$^{-1}$ on day 10 
(Pastorello et al. 2008) and 
photospheric radius $r\sim vt\sim10^{15}$ cm the effective
temperature of the continuum turns out to be $\sim10^4$ K.

A question arises, whether a helium-rich CDS with a mass of
 $\sim0.05~M_{\odot}$ is able to maintain a smooth continuum at the 
age of $\sim10$ days?
To illustrate the possibility I consider the radiation transfer in the 
isothermal helium slab with Thomson 
scattering and absorption taken into account. 
To facilitate  estimates I adopt the LTE approximation. With 
the boundary condition 
$F(0)=cU(0)/\sqrt3$ ($U$ is the radiation density) is then
\begin{equation}
F_{\nu}=\frac{4\pi}{\sqrt3} B_{\nu}(T)\frac{\sqrt\epsilon(1-E)}
{1+E+\sqrt\epsilon(1-E)}\,,
\label{eq-slab}
\end{equation}
where $E=\exp(-\tau\sqrt{3\epsilon})$, $\tau$ is the extinction
optical depth and $\epsilon$ is the thermalization parameter, i.e.,
 absorption to extinction ratio; both $\tau$ and 
$\epsilon$ are functions of frequency. This expression permits us to 
calculate luminosity of the CDS for a certain gas temperature 
using Kramers-Unzold approximation
for the absorption coefficient of the helium-rich matter. 

The density in the CDS is found from the pressure equilibrium 
adopting the  wind density parameter
$w=4\times10^{16}$ g cm$^{-1}$, shell velocity of $10^4$ km s$^{-1}$, 
and the radius of $\sim10^{15}$ cm. 
To produce the continuum luminosity of $\sim10^{43}$ erg s$^{-1}$ 
the equation (\ref{eq-slab}) requires the gas of $\sim12500$ K for the 
CDS mass of $0.05~M_{\odot}$. 
In the range $\lambda>4000$ \AA\ the model spectrum is close to black-body 
with the temperature $T\sim10200$ K. 
Although the extinction optical depth of the shell at 6000~\AA\ is 
only $\sim0.4$ the equilibration is rather
efficient because thermalization parameter is large $\epsilon>0.92$ in 
the range $\lambda>3000$~\AA.
We thus conclude that the smooth continuum may arise in early SN~Ibn. 
The epoch of the early smooth continuum of SN~2006jc preceeded the 
discovery and therefore was missing.

\subsection{Blue bumpy continuum at $t\geq20$ d}
\label{sec-gen3}

The smooth continuum of SN 2000er grew bumpy 
on a time scale of about one week (Pastorello et al. 2008), 
although the short period of SN~2000er observations does not permit one 
to follow 
the early evolution from smooth till fully developed bumpy continuum we 
see in SN~2006jc on day 22. 
No doubt, the bumpy continuum is composed by numerous lines, 
presumably of Fe\,II as suggested by Foley et al. (2007). However, unlike any 
normal SN~Ibc, including SN~1998bw, the spectum of SN~2006jc
is unusually blue: the colour $B-R=-0.45$ (Foley et al. 2007)
compared with 0.48 of SN~1998bw at the light maximum (Galama et al. 1998).
The spectrum also does not show broad absorption lines characteristic 
of SN~Ibc. This indicates that the configuration of "photospere" 
surrounded by an extended free expansion atmosphere is irrelevant 
in this case. 
Following the picture invoked for the early smooth continuum 
I assume that the bumpy blue continuum of SN~2006jc arises from the 
same CDS but this time with the dominant emission in metal lines. 
This picture is similar to SN~2002ic (Chugai et al. 2004a), so 
to demonstrate the possibility I use the same model and the same line list 
of $\sim18000$ lines of iron-peak elements in the range of 3500-10000 \AA.

The calculated continuum and observed SN~2006jc spectrum on day 22 
(Foley at al. 2007) are shown in Fig. \ref{f-synt}.
The model suggests the emission of randomly oriented plane sheets 
distributed in the shell $\Delta R/R=0.1$. The total mass of cold fragments 
is $0.05~M_{\odot}$ and the  solar mass fractions of iron peak elements 
(Ti, V, Cr, Fe, Ni) are assumed. The adopted 
expansion velocity is $9000$ km s$^{-1}$, ionization temperature 7000 K, 
excitation temeperature 8000 K, and the area ratio $A=S/4\pi R^2=20$. 
Just to recall, the area ratio is the ratio of cumulative surface area 
of fragments to the surface area of the CDS; this parameter 
 characterizes a mixing degree. 
 The model spectrum is sensitive to the variation of the 
  excitation temeperature within 10\% and in lesser extent to 
  to the shell mass and thickness, which may vary within 30\% without 
  pronounced change of the spectrum.
The adopted value of the shell thickness $0.1R$ corresponds approximately 
to the width of mixing layer formed as a result of a 
Rayleigh-Taylor instability of the thin shell (Chevalier \& Blondin 1995).
 The model qualitatively reproduces the observed spectrum and bolometric 
luminosity $\sim3\times10^{42}$ erg s$^{-1}$ 
(cf. Mattila et al. 2008). The high relative intensity of the blue continuum 
is the result of a larger number of lines per unit wavelength interval 
in the blue compared to the red part 
of the spectrum and a weak saturation of intensity in the blue 
for a moderate area ratio. 
Given satisfactory fit and sensible values of model parameters, the proposed 
picture for the blue continuum provides us indirect support for the 
conjecture of the strong early CS interaction in SNe~Ibn.

\begin{figure}
\includegraphics[width=80mm]{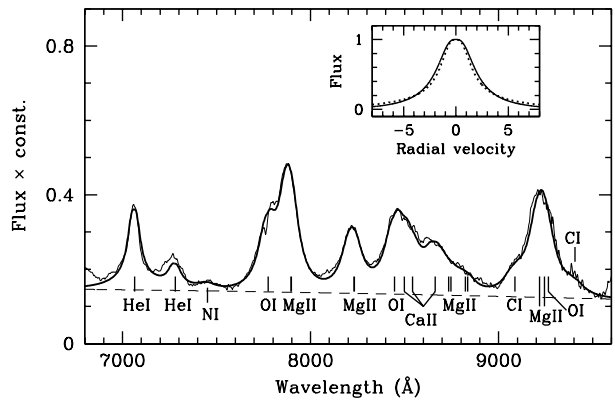}
\caption{
Emission lines in the red part of the spectrum of SN~2006jc 
on day 22. Thick line shows synthetic spectrum; 
thin lines is the observed spectrum (Fooley et al. 2008); dashed line 
is adopted continuum. Inset shows 
profiles used in the synthetic spectrum for He\,I lines (dotted line) 
and metal lines (solid line). 
}
\label{f-mg}
\end{figure}

\subsection{CS interaction and emission lines}
\label{sec-gen4}

Spectra of SN~2006jc (Foley at al. 2007; 
Pastorello et al. 2007) suggest that 
the velocity of He\,I line-emitting gas lies in the range of
$800-6000$ km s$^{-1}$ with the velocity of the bulk of emitting material 
in the range of $800-3000$ km s$^{-1}$. The minimal velocity of 
$\sim 800$ km s$^{-1}$ could be associated with the 
cloud shocks speed and perhaps with the velocity of the undisturbed CSM. The 
latter is about $620$ km s$^{-1}$ as indicated by the absorption 
minimum of narrow lines of 
O\,I 7773 \AA\ and He\,I 6678 \AA\ (Anupama et al. 2008). 
Interestingly, the expansion velocity of CSM of 
SN~2000er indicated by narrow He\,I absorption components 
in early spectra is 800-900 km s$^{-1}$ (Pastorello et al. 2008), 
close to this value. 

To account for the velocity spectrum of the He\,I line-emitting gas 
I suggest that the bulk of the  He\,I emission originates 
from shocked CS clouds.
Two-dimensional hydrodynamic simulations of 
interstellar cloud engulfed by the blast wave 
(Klein et al. 1994) prompt us a picture in which 
the velocities of the line-emitting gas are produced 
by slow (because of density contrast) radiative shocks of CS clouds 
in combination with 
the fragmentation cascade of shocked clouds and acceleration of fragments.
This scenario is similar to that proposed for the H$\alpha$ emission 
in SN~2002ic (Chugai et al. 2004a). 
The total optical luminosity of He\,I lines is about 
$L\sim5\times10^{40}$ erg s$^{-1}$ on day 22 (Smith et al. 2008). 
The corresponding radiation efficiency of the forward shock kinetic 
luminosity is $\eta=L/(0.5wv_s^3)$. 
With the forward shock velocity $v_s=9000$ km s$^{-1}$, 
wind density parameter $w=4\times10^{16}$ g cm$^{-1}$ the required 
efficiency should be $\eta\sim0.003$, a reasonable value.

The other emission lines seem to originate from the same line-emitting 
gas. This is demonstarted by the synthetic spectrum of SN~2006jc 
on day 22 in which  
profiles and intensities were adjusted to fit the observed spectrum
(Fig. \ref{f-mg}).
Apart from He\,I, O\,I, and Ca\,II lines the synthetic spectrum 
includes Mg\,II lines (Pastorello et al. 2007; Anupama et al. 2008) 
and two C\,I lines 9088 \AA\ 
 ($^3$P$^o$--$^3$P) and 9406 \AA\  ($^1$P$^o$--$^1$D), which are 
 also pronounced in novae (Williams et al. 1991).
There is a vague hint of a week N\,I 7452 \AA\ emission 
 with the intensity ratio  N\,I 7452 \AA/O\,I 7773 \AA\ of $\sim0.1$ 
(Fig. \ref{f-mg}). 
The fact that profiles of all the lines are identical 
confirms that these emissions originate from the 
same line-emitting gas, which I associate with the shocked CS clouds. 

\begin{figure}
\includegraphics[width=80mm]{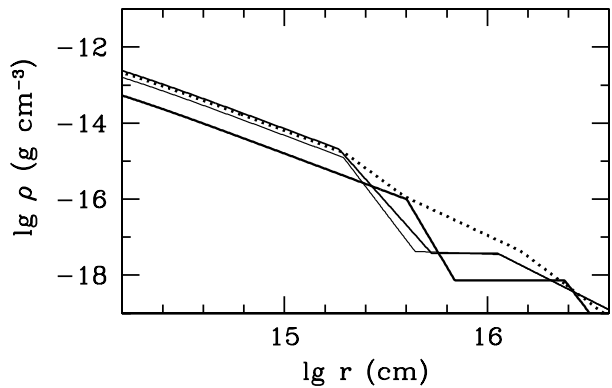}
\caption{
Density distribution of the circumstellar gas for studied models. 
Thin, intermediate and thick solid lines correspond to the model A1, 
A3, and A2 
respectively, while the dotted line stands for the model B.
}
\label{f-den}
\end{figure}

\section{Interaction model}
\label{sec-interact}

The CS interaction and bolometric light 
curve are simulated using the model described earlier (Chugai 2001).
To recap, the hydrodynamics of the interaction
 is treated in the thin shell approximation (Chevalier 1982a). 
The X-ray emissions of the reverse and forward shocks are calculated 
assuming a homogeneity of each post-shock layer and adopting 
the cooling function of Sutherland \& Dopita (1993). The electron 
temperature in the postshock 
gas is calculated taking into account partial Coulomb equilibration.
The absorbed and emergent X-ray luminosity is computed assuming the 
unabsorbed emissivity spectrum to be 
 $\epsilon (E) \propto E^{-0.5}\exp(-E/kT)$ and taking into acount the 
absorption by the cool gas (SN ejecta, CDS and unshocked CSM). 
The model despite many simplifying assumptions seems to be 
reasonable which is confirmed by the modelling of SN~1999em X-ray flux 
(Chugai et al. 2007). 
The absorbed X-ray luminosity is reprocessed into the
ultraviolet (UV), optical, and infrared photons. We adopt that 
the bulk of this radiation falls in the band in which the observed 
bolometric light curve is sampled. This assumption is sensible at the 
early epoch ($t<1$ month), when the cool gas is opaque in UV region
and rather crude at the late epoch ($t\sim 1$ yr), when significant 
fraction of the reprocessed radiation can escape in UV lines 
(Chevalier \& Fransson 1994). The interaction luminosity  combined with 
the diffusion luminosity of the ejecta produces the model bolometric
luminosity.
In turn, the diffusion luminosity is modelled using the Arnett (1980) 
approximation. 

A smooth CSM is assumed, neglecting clumpiness, which implies that 
the calculated luminosity and temperature of the X-ray emission from the
forward shock perhaps are not reliable. I present two versions of the 
interaction model: with $^{56}$Ni and without $^{56}$Ni.
The gamma-ray deposition is calculated assuming that $^{56}$Ni is mixed 
in the inner 90\% of the ejecta mass.
The ejecta density distribution is approximated by the 
exponential law $\rho\propto \exp(-v/v_0)$, while the radial density 
distribution of the CSM is adjusted to describe the bolometric 
 and X-ray luminosities
and the radii of the dusty shell associated with the CDS.
The composition of major elements is assumed to be 
He : C : O = 0.9 : 0.06 : 0.04. This choice, called a "standard abundance", 
is similar to that of Mattila et al. (2008).
The pre-SN radius $R_0=10~R_{\odot}$ is assumed; the exact value 
is irrelevant, unless the radius is very large, $R_0\gg10~R_{\odot}$. 

The Table 1 shows  model parameters: the explosion energy, 
ejecta mass, $^{56}$Ni mass, 
mass of the CS envelope, mass of the CDS formed in the reverse shock 
by day 50, and the wind density 
parameter $w$ in the inner region $r<2\times10^{15}$ cm. 
Our models admit formation of the CDS with a mass of 
$0.01-0.09~M_{\odot}$ in the forward 
shock during initial phase of 5-15 days depending on the model.
In a more realistic case of clumpy CSM the situation with the forward shock 
CDS would become more complicated. Radiative cloud shocks would 
result in the formation of cool CS gas in a forward shock.
However, the expected kinematics and distribution of this cool gas 
would be essentially different compared to the model of a smooth wind.
Models with $^{56}$Ni differ by the ejecta parameters and the wind density. 
It should be stressed that ejecta mass and energy are poorly constrained 
by the light curve and spectrum. The high energy model A2 
corresponds to the preferred model of Tominaga et al. (2008), while 
the models A1 and A3 have moderate explosion energy. 
The CSM density distributions for all the models
are presented in Fig. \ref{f-den}. 

\begin{figure}
\includegraphics[width=80mm]{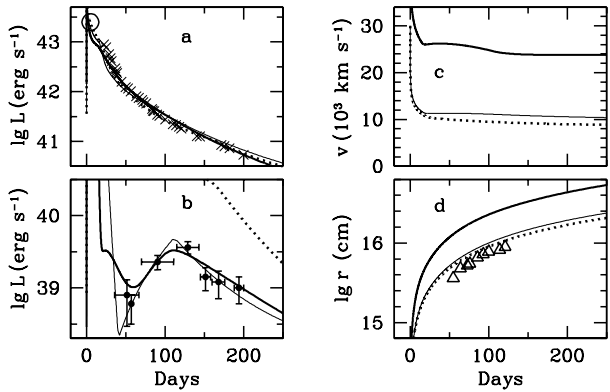}
\caption{
Left panels: Bolometric light curves ($a$) and X-ray luminosity ($b$) 
escaping after absorption in unshocked ejecta, CDS and CSM for 
the models A1 (thin solid line), A2 (thick solid line) and B (dotted line). 
Crosses show the observed bolometric light curve according to 
Pastorello et al. (2008);  circle represents the luminosity of SN~1999cq 
at the light maximum; squares show X-ray luminosity in 0.2-10 keV band 
(Immler et al. 2008).  
Right panels: Velocity ($c$) and radius  ($d$) of the cool dense shell.
Triangles show radii of the dusty shell reported by Mattila et al. (2008).
}
\label{f-dyn}
\end{figure}

All the models satisfactorily reproduce the observed bolometric light curve 
(Fig. \ref{f-dyn}a) including the maximal luminosity of SN~1999cq 
estimated from  $M_R$ magnitude (Pastorello et al. 2008).
Note, we do not plot the low-mass model A3 since its 
behavior repeats that of the model A1 in all the panels of Fig. \ref{f-dyn}.
Models with $^{56}$Ni (A1 and A3) reproduce the X-ray flux within 
$1\sigma$ errors (Fig. \ref{f-dyn}b), while the model A2 with the high 
energy lies beyond $1\sigma$ range. The model B
predicts too high X-ray 
luminosity that exceeds by a factor of $\sim5\times10^2$ the observed 
X-ray luminosity 
around day 50. This is a serious problem for the model without $^{56}$Ni. 
Yet one cannot rule out that 
the clumpiness of CSM, ignored here, might modify the interaction 
zone in such a way that X-rays were efficiently absorbed. 
The observed X-ray flux seem to show a drop by a factor 2.5 between the 
maximum on day $\approx110$ and the epoch of $\approx135$ d 
(Immler et al. 2008).
This drop cannot be reproduced in any of our interaction model, 
 although models A1 and A3 are consistent with observations 
within $1\sigma$ errors. If the drop is real then shocks 
in CS clouds engulfed by the forward shock layer contribute 
significantly to the X-ray emission. In this case the drop of X-ray flux 
could be related with the disappearence of CS clouds beyond the radius
$r\sim10^{16}$ cm.

A massive ($0.2-0.3~M_{\odot}$) CDS forms by day 50
in all the models (Table 1); the expansion velocity at the age 
of 50-100 days is $\sim10^4$ km s$^{-1}$ in models A1, A3 and B 
and $\sim26000$ km s$^{-1}$ in the high-energy model A2 (Fig. \ref{f-dyn}c).
Accordingly, for models A1, A3 and B the CDS radii are close to 
photometric radii 
of the dusty shell (Fig. \ref{f-dyn}d) reported by Mattila et al. (2008), 
while the high-energy model A2 has essentially larger CDS radius. 
Interestingly, for the considered models the CDS mass at $t=5$ d falls 
in the range $0.05-0.06~M_{\odot}$, the value used in Section 2 to analyse 
the smooth continuum of SN~2000er.

I explored also the model in which the CSM with the mass of 
$0.1-0.5~M_{\odot}$ has a structure of a dense thin shell with the 
radius of $4\times10^{15}$ cm. This choice corresponds to
a scenario in which the CS shell 
was violently ejected with the velocity of $600$ km s$^{-1}$ 
during the outburst two years prior to the SN explosion.
In this case the bolometric light curve is notably affected by 
the CS interaction and the model light curve with or without $^{56}$Ni 
is inconsistent with observations.

To summarize, the bolometric light curve of SN~2006jc can be explained by 
models with and without $^{56}$Ni; both cases suggest the dense wind 
at the distance $r\leq2\times10^{15}$ cm. In all the models massive CDS 
forms which might become the dust formation site.

\begin{table}
  \caption{Parameters of models of SN ejecta and CSM }
  \begin{tabular}{ccccccc}
\hline
Model & $E$ & $ M$ &  $M_{Ni}$ & $M_{cs}$ &
$M_{cds}$ & w \\ 
   &   $10^{51}$ erg & \multicolumn{4}{c}{$M_{\odot}$} &
   $10^{16}$ g/cm \\ 
\hline

 A1 & 1.5   &  3    &  0.12  & 0.14   & 0.21  & 6 \\
 A2 & 10    &  5    &  0.2   & 0.08   & 0.19  & 2\\ 
 A3 & 1     &  1    &  0.3   & 0.18   & 0.21  & 9 \\
 B & 1.5    &  3    &  0     & 0.27   & 0.28  & 8 \\
 
\hline
\end{tabular}
\label{t-dmod} 
\end{table} 

\section{Dust formation}
\label{sec-dust}

For the dust to condense the gas and dust temperatures should 
become lower than the condensation temperature ($T_c$). 
Unfortunately, the CDS gas temeperature ($T_g$) cannot be reliably 
determined from the radiative energy balance because of complexities 
of molecular compounds and processes involved. We therefore adopt a sensible 
assumption that $T_g$ 
is equal to the radiation temperature  $T_r=(F/ac)^{1/4}$,  
where $F$ is the radiation flux at the CDS radius, $a$ is the radiation 
 constant, $c$ is the speed of light. This approximation probably is not 
perfect because the gas irradiated by diluted 
hot radiation is usually warmer than at the radiation temperature. 
To take into account a possible deviation from the radiation temeperature 
I also consider the case, in which the 
gas temperature is equal to the dust temperature computed in a formal way; 
this approximation
presumably crudely reflects the physics of heating and cooling of molecular gas.
The dust temperature $T_d$ is calculated 
from the balance between the absorption of SN radiation and the dust emission. 
We approximate the SN radiation by the diluted black body spectrum with 
the temperature 
of $10^4$ K; dust grains are assumed to be graphite spheres with the radius 
of 0.03 $\mu$m and the absorption efficiency is set according to 
Draine \& Lee (1984). Calculations show that on day 50 
the dust temperature turns out $\sim10$\% larger than the 
radiation temperature.

Given $T_g<T_c$, the dust, nevertheless, forms only, if a characteristic 
time of the dust formation 
is essentially smaller than the expansion time. 
The time scale of the dust formation is determined by 
the nucleation rate and growth of nuclei due to the gas 
accretion onto the grains. 
The rate of the dust growth can be easily estimated using results 
of the interaction modelling. 
At the age $t=50$ d the carbon number density in the CDS
for the explored models with standard abundance 
is $n\sim(2-10)\times10^{11}$ cm$^{-3}$ 
assuming that all the oxygen is bound in CO.
Adopting $T_g=1800$~K and sticking parameter of 0.5 we find 
that it takes $\sim(1-3)\times10^2$~s  for the dust grains to grow 
$a\sim10^{-5}$ cm. This shows that the duration of the 
dust condensation probably is not constrained by the dust growth.

In line with the theory of homogeneous nucleation (cf. Drain \& Salpeter 1977;
Hasegawa \& Kozasa 1988), the nucleation time scale can be 
expressed via the number density of dust grains $n_g$ of the final 
radius $a$ and nucleation rate $J$
\begin{equation}
t_{nuc}=n_g/J\,,
\label{eq-nuctime}
\end{equation}  
where $n_g=n\Omega/(4\pi a^3/3)$ and
 $\Omega\sim 2\times10^{-23}$ cm$^{-3}$ 
is the volume per carbon atom in a condensed phase.
The nucleation rate (Draine \& Salpeter 1977; Hasegawa \& Kozasa 1988) is 
defined by the accreation rate onto clusters close to equilibrium point
\begin{equation}
J=\alpha4\pi r_c^2uZn^2\exp(-16\pi\sigma^3/3g^2kT)\,, 
\label{eq-nucr}
\end{equation}  
where $\alpha$ is the sticking parameter (we adopt $\alpha=0.5$), 
Zeldovich factor $Z$ takes into account 
non-equilibrium distribution of nuclei sizes,
$u$ is the average thermal velocity of carbon atoms, $r_c=-2\sigma/g$ 
is the radius of a cluster of critical size, 
$\sigma=1400$ erg cm$^{-2}$ is the energy of graphite surface tension
(Tabak et al. 1975), $g$ is the change of free energy in the transition 
from gas to condensed phase
\begin{equation}
g=-(kT/\Omega)\ln(P/P_{sat})\,,
\label{eq-genergy}
\end{equation}  
where $P$ and $P_{sat}$ are the partial pressure and 
saturation pressure of carbon. 

\begin{figure}
\includegraphics[width=80mm]{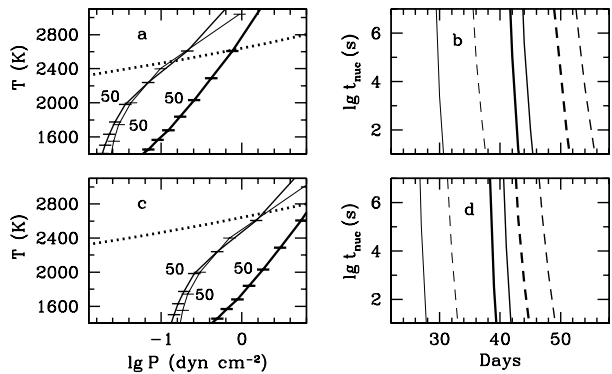}
\caption{
Left panels: Loci of the radiation temperature and pressure in the CDS 
with the standard composition (upper panel) and five times larger carbon 
abundance (lower panel). Thick solid line is for model B, 
intermediate line is for model A2 and thin line is for model A3. 
Tick marks correspond to every 5 days; the tick on day 50 is indicated. 
The phase equilibrium for graphite (dotted line) is taken according 
to Haines \& Tsai (2002).
Right panel: Evolution of nucleation time in the CDS with the 
standard composition (upper panel) and five times larger carbon 
abundance (lower panel). Cases of the gas temperature  equal to the radiation 
and the dust temperature are shown as solid and dashed lines
respectively.
The line thickness increases along the sequence  A3, A2, B.
}
\label{f-nuc}
\end{figure}

The evolution of partial pressure of carbon and gas temperature 
for models A2, A3, and B is shown in Fig. \ref{f-nuc}a. 
For the model A1, which is not shown, the behavior of calculated 
values is intermediate between A2 and B.
Using a standard expression for Zeldovich factor (cf. Hasegawa \& Kozasa 1988) 
and the partial pressure of carbon from the interaction models 
one can compute from equations (2-4) the nucleation time 
(Fig. \ref{f-nuc}b) for the adopted grain 
 final radius $a=10^{-5}$ cm. It should be stressed that 
 the result is not sensitive to adopted $a$. For $a=10^{-6}$ cm the 
 value $t_{nuc}=10^4$ s is attained by $0.8$ day later. 
Apart from the standard mixture, we explored also the case of five times
 higher carbon and oxygen abundance 
  He : C : O = 0.5 : 0.3 : 0.2 (Fig. \ref{f-nuc}c,d). In this case 
the nucleation occures a bit earlier.  

A steep time dependence of $t_{nuc}(t)$ function permits us 
to recover the dust formation epoch for each particular model. 
Defining the dust formation epoch by the condition 
$t_{nuc}=1$ day we find for the standard abundance that dust 
forms between days 42 and 54 for the low energy models and 
at 30-36 d for the high energy model A2. 
In the case of five times higher carbon abundance the dust forms 
at 40-48 d for the low energy models and at 28-32 d for the high energy 
model. The earlier dust-forming phase in the high energy model 
is related to the higher CDS velocity, i.e., lower gas temperature 
at the same age.
This modelling shows that the dust forms in the CDS with the standard 
abundances of He, C, and O at the epoch close to day 50. 

The maximal amount of dust for the standard abundance 
is $\sim0.005~M_{\odot}$ assuming that all the oxygen is bound in 
CO molecules and the remaining carbon is converted into the dust.
Around day 100 the CDS with the radius of $\sim10^{16}$ cm 
(Fig. \ref{f-dyn}) and the dust mass of $\sim0.005~M_{\odot}$ 
has the optical depth  $\tau\sim500$ in $R$ band
assuming extinction efficiency $Q_e=2\pi a/\lambda$. 
The found upper limit is by a factor of $\sim10^2$ larger than observational 
estimates (Mattila et al. 2008) which indicates either low efficiency of the
dust formation, or clumpy structure of the dusty shell.
Another interesting point is that the one-zone model considered above 
predicts rapid dust condensation,
which is not the case: observations show slow increase of the amount of 
dust in SN~2006jc (Mattila et al. 2008). This means that conditions 
in the CDS matter perhaps are not uniform. A spacial variation of the 
gas temperature is probably
a primary factor responsible for the gradual process of the dust condensation. 

We considered above the dust formation only in the CDS created by 
the reverse shock. However, one cannot rule out that 
the cool gas of shocked CS clouds in the forward shock 
might be also an appropriate site of 
the dust formation as well. With some modification this possibility 
corresponds to the scenario of the dust formation in the cool shell 
of the forward shock discussed by Mattila et al. (2008).

\section{Dusty shell and blue shift}
\label{sec-bshift}

The scenario proposed here suggests that He\,I emission lines
 originate from fragmented shocked CS clouds in the forward 
shock. To simulate line profiles affected by the dust absorption one need to 
specify the distribution of emissivity and velocity of the line-emitting gas.
I suggest that the unabsorbed line profile is determined by the 
flow of cool fragments of shocked CS clouds in the range $R_1<r<R_2$, 
where $R_1=1$ is the radius of the CDS, or more precisely, 
of the contact surface between shocked ejecta 
and shocked CS gas, and $R_2$ is the radius of the forward shock front 
adopted to be 1.25 (cf. Chevalier 1982b). 
A possible collision of a cloud with the CDS would result in the sudden 
acceleration of the cloud shock; the shock gets adiabatic and cloud 
is demolished instantly. The resulting fragments in this case get 
hot and do not contribute in He\,I emission.
The integrated emissivity is assumed to be constant 
in the range $R_1<r<R_2$, which  
is sensible approximation, if the bulk of CS clouds crosses 
the postshock layer before they get completely fragmented. 

The velocity distribution of the line-emitting gas generated by 
CS clouds interaction with the forward shock is not trivial.
The shocked cloud acquires a cometary structure
with the low velocity core and high velocity tail of fragments 
(Klein et al. 1994). For particular cloud 
the flow kinematics is approximately monotonic: the velocity increases 
from the core towards farmost tail. The smallest fragments have maximal 
velocities but they also are prone to rapid mixing with the 
hot gas (Klein et al. 1994). We assume that the smallest fragments of 
cool gas dissapear before they acquire the velocity of the forward 
postshock flow. From the mass conservation it follows that 
the average density of cool gas in the tail monotonically decreases with 
velocity. The combined flow of clouds 
at different stages of interaction cannot be reduced 
to a single-valued function $v(r)$. Instead, 
at the given radius a broad spectrum of velocities $f(v)$ is expected 
with the minimal velocity corresponding to cloud shock speed 
$v_1\sim10^3$ km s$^{-1}$ and maximal velocity
$v_m(r)$ monotonically rising with $r$ and determined by 
"oldest" shocked clouds that are colliding with 
the CDS. I adopt the linear law for the maximal velocity
$v_m=v_1+(v_2-v_1)(r-R_1)$, where $v_2$ is a fitting parameter.
The velocity density distribution function at the given radius 
is taken in the simple form 
\begin{equation}
f(v)=C\left(\frac{v_m-v}{v_2-v_1}\right)\,, 
\end{equation}
where $C$ is normalizing factor.

The CDS is subject to Rayleigh-Taylor 
instability, fragmentation, and mixing in the forward shock layer
(Chevalier 1982a; Blondin \& Ellison 2001), so the dusty CDS material 
should be spread between the contact 
surface $R_{d,1}=R_1=1$ and some external radius $R_{d,2}$ which is a free 
parameter.
The emission line profiles affected by the dust is simulated here 
using Monte-Carlo technique. The dust in our model has velocity 
of the CDS for which we adopt 9000 km s$^{-1}$. This parameter 
weakly affects the scattered radiation. The Rayleigh 
phase function and scattering albedo $\omega=0.6$ are assumed.
I checked different values of $\omega$ and 
found only minor differences which refer to 
the intensity of the broad scattered component. Note, this component, 
although weak, may be revealed in the polarized light, if the 
asymmetry of dust distribution is significant.
I also tried the phase function dominated by back scatter which 
would be the case, if the dust resides in optically thick clumps. 
The difference, primarily in scattered component, turns out small.

\begin{figure}
\includegraphics[width=80mm]{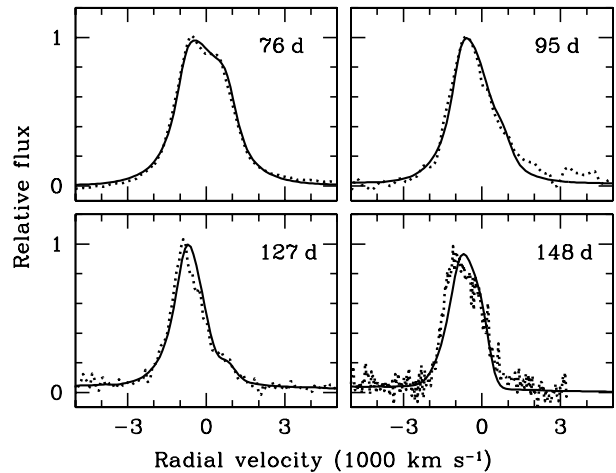}
\caption{
Line profile of He\,I 7064 \AA\ in SN~2006jc spectra for four epochs. 
The model and observations (Foley et al. 2008) are shown 
as a solid line and dotted line, respectively.
}
\label{f-ocu}
\end{figure}

The optimal model profiles 
are plotted (Fig. \ref{f-ocu}) together with 
observed He\,I 7065 \AA\ line profile of SN~2006jc for different epochs 
(Foley at al. 2007). The fit is good for the three first dates and 
sensible in the last plot. This can be considered as success of 
the proposed model. The recovered parameter $v_2$,
extinction optical depth and radii of the dusty shell are given in Table 2. 
The decrease of $v_2$ suggests that the contribution 
of the high velocity line-emitting gas falls with time. Note, the 
parameter $v_2$ is determined by unabsorbed blue part of profile and does not 
depend on the dust distribution and the optical depth.
It is not clear whether the decrease of the
high-velocity component is related with the 
cloud fragmentation and fragment acceleration or with the 
mixing of high velocity fragments in the hot gas. 
There is a hint that the relative radius of the 
outer boundary of the dusty shell $R_{d,2}$ increases with time. 
 Interestingly, such a behavior qualitatively agrees with 
the initial evolution of a mixing layer driven by the
Rayleigh-Taylor instability (Chevalier \& Blondin 1995; 
Blondin \& Ellison 2001). 
A detailed comparison with published 
numerical models is, however, not possible, 
because the models for SN~2006jc presented here do not have 
power-law density profiles for the ejecta and the wind.

I explored cases with larger inner radius of the dusty shell, 
$R_{d,1}>1$, but found that the fit worsens  
in this case. This means that in the preferred model the center of 
the dust layer is shifted inward 
relative to the center of the line-emitting layer. This is expected 
in the proposed picture in which most dust originates from 
the fragmented CDS and the line-emitting gas is related with 
the shocked CS clouds. 

\begin{table}
  \caption{Parameters of the line profile model}
  \begin{tabular}{cccccc}
\hline
 Day   & $v_2$    & $\tau$ & $R_{d,1}$ & $R_{d,2}$ \\
   & km s$^{-1}$ &        &          &         \\
\hline
  76     &  6000       & 0.07    &    1  &  1.1\\
  95     &  5000       & 0.4     &    1  &  1.1\\ 
  127    &  4000       & 1.1     &    1  &  1.2\\ 
  148    &  3000       & 5       &    1  &  1.2\\
\hline
\end{tabular}
\label{t-prof} 
\end{table} 

\section{Discussion}
\label{sec-discuss}

The proposed scenario of SN~Ibn with the strong early CS interaction 
turns out successful 
in explaining some unusual observational properties of SNe~Ibn. However, 
until now we ignored some questions that might have 
interesting consequences for our model. It is worthy, therefore, to 
pose and briefly discuss challenging issues and, if possible, to 
impose additional constraints for the model.

\subsection{What does power blue continuum?}

The proposed interpretation of the blue bumpy continuum suggests 
that the most emergent optical radiation originates from the CDS. 
In case, when the interaction luminosity dominates, the energy 
from the reverse and forward shocks is deposited into the CDS in a usual way 
via absorption of X-rays and, perhaps, the thermal conductivity in the 
mixing zone of the forward shock.

Another possibility could be that the dominant source is $^{56}$Ni decay. 
In this case the energy is deposited  into CDS 
owing to the Compton scattering of gamma-quanta 
and the absorption of ultraviolet radiation emitted by the unshocked ejecta 
in turn powered by the radioactive decay. 
The efficiency of the gamma-ray deposition to the CDS is of the order 
of the ratio of the CDS mass to the mass of ejecta, which is small 
unless ejecta mass is low $\leq1~M_{\odot}$.

Alternatively, the CDS gas could be ionized and excited 
by the absorption of the ultraviolet radiation from the unshocked ejecta. 
The problem with this mechanism is that in the unshocked freely expanding 
SN~Ibc envelope the ultraviolet radiation is strongly reprocessed by 
the photon scattering and splitting in numerous ultraviolet metal lines,  
so the emergent ultraviolet flux gets strongly suppressed relative to 
visual. The situation in this respect is similar to SNe~Ia 
(cf. Pinto \$ Eastman 2000).
At first glance the ultraviolet flux might become stronger,  
if the ejecta mass is low. However, for this to be the case the 
mass should be much lower than that of SNe~Ia, i.e., $M<1~M_{\odot}$. 
We thus conclude that in the radioactive model 
the CDS can be the source of the blue continuum, if the ejecta 
mass is low, $\leq1~M_{\odot}$.

\subsection{Where does He\,I absorption form?}

Helium lines (e.g. 3889 \AA) in the blue part of SN~2006jc spectrum 
show P Cygni profiles with the absorption minimum 
at $\sim-1000$ km s$^{-1}$ and a comparable equivalent width of 
absorption and emission components (Foley at al. 2007).  
A natural explanation of this phenomenon is 
that the continuum in blue region is strong, so the 
scattering component dominate or comparable with the net line emission.
In fact absorption components are also seen in O\,I 7773 \AA\ and 
in He\,I 6678 \AA\ lines (Anupama et al. 2008). 
The maximal velocity observed in the blue absorption wing of He\,I 3889 \AA\ 
line is 
$\approx 1900$ km s$^{-1}$, three times larger than 
the wind velocity, $v\approx 600$ km s$^{-1}$. 
Apart from the wind, therefore, an absorbing high velocity component
($>600$ km s$^{-1}$) is needed to account for the line absorption. 

Three possibilities are conceivable for the line-absorbing 
gas in the range of 600-2000 km s$^{-1}$: 
(a) the CSM in fact has the free expansion kinematics $v\propto r$ with 
the most material in the range of 600-2000 km s$^{-1}$;
(b) the line-absorbing gas is simply the same line-emitting gas 
in the forward shock; 
(c) the CSM is flow with the constant velocity of $\sim600$ km s$^{-1}$ 
but in the preshock zone the CS gas is accelerated up to 1900 km s$^{-1}$. 
The first option predicts that the emission lines related with the 
CSM should get broader with time. In fact, the absorption of 
He\,I 3889 \AA\ does not show this trend according to spectra 
taken by Foley et al. (2007).
The second option suggests that fragments of CS clouds in the forward shock 
should have large covering factor $\sim1$, which is 
unlikely, although is not rulled out.
The third possibility is similar to that suggested for the H$\alpha$ 
high velocity CS absorption component in early SN~1998S (Chugai et al. 2001), 
where this feature was attributed to the preshock gas accelerated
 by either radiation or cosmic rays generated in the forward shock.
At the distance of $\sim10^{16}$ cm the radiative force in case of SN~2006jc 
is weak and cannot accelerate the preshock gas up to more than 
100 km s$^{-1}$ in 
the wind frame and this mechanism should be rejected.

As to cosmic rays, the situation is more optimistic. 
The point is that in case of clumpy CSM the cosmic ray pressure is build up 
by the overall shock kinetic luminosity, while 
the acceleration is applied primarily only to the intercloud gas 
with a small mass fraction of the inflowing CSM. The intercloud gas, 
therefore, can acquire significant velocity in the preshock zone.
Let $\rho_{ic}$ is the density of the intercloud gas that is 
substantially lower than the average CS density $\rho$. 
Momentum and mass conservation result in the 
maximal preshock velocity of the intercloud gas 
\begin{equation}
u=\frac{1}{6}\phi v_s\frac{\rho}{\rho_{ic}}\,,
\end{equation}
where $v_s$ is the forward shock speed, $\phi$ is the 
ratio of the energy density of cosmic rays to the density 
of the shock kinetic energy  $0.5\rho v_s^2$.
For $v_s=9000$ km s$^{-1}$ and $\rho/\rho_{ic}=5$ one needs 
$\phi\approx0.2$ to produce the preshock velocity of 
$u\approx1300$ km s$^{-1}$ in the wind frame.
This estimate demonstrates that velocities of absorbing gas 
up to 1900 km $^{-1}$ might be explained, if the line-absorbing gas is 
associated with the intercloud gas accelerated in the cosmic 
ray precursor.

\subsection{Why intensities of Mg\,II and O\,I lines are similar?}
\label{sec-mg2ando2}

Between days 22 and 60 intensities of Mg\,II 7889 \AA\ and O\,I 7773 \AA\ 
lines in SN~2006jc are comparable (cf. Anupama et al. 2008).  
Given similar ionization and excitation potentials
of these species a straightforward implication of this fact is 
that the abundances of oxygen and magnesium are comparable. This 
indicates that 
we perhaps see the matter in which oxygen is strongly depleted by CNO burning. 
Indeed, massive stars at the boundary between helium core and hydrogen 
envelope have a zone, in which oxygen is depleted owing to CNO burning 
by a factor of 10-20 (cf. Hirschi et al. 2004; Meynet \& Maeder 2003). 
The low abundance of hydrogen in the CSM around SN~2006jc is consistent 
with the conjecture that we might see the matter with strongly 
depleted oxygen. 

An alternative explanation is that Mg/O ratio is cosmic, and 
intensities of Mg\,II and O\,I lines are comparable because these lines are 
thermalized, i.e., level populations are close 
to Boltzmann distribution and lines are optically thick. 
The thermalization in a certain line occurs if
$\epsilon_{21}N_s>1$, where $N_s$ is the average number of conservative 
scattering in the line before the photon escape,
$\epsilon_{21}=q_{21}n_e/A_{21}$ is the thermalization parameter,
 and $q_{21}$ is the collisional de-excitation rate. 
Using estimates based on the CSM density,  
pressure equilibrium, He\,I line intensity, and Boltzmann excitation 
of the lower level I estimate $\epsilon_{21}N_s\sim10$ for 
the Mg\,II 7889 \AA\ line on day 22. This confirms the 
possibility of the thermalization of Mg\,II and O\,I emission lines.

What explanation of the comparable intensities of  Mg\,II 
and O\,I lines --- similar abundances of oxygen and magnesium or 
thermalization --- is correct remains an open question. 
The conjecture of the similar abundances of Mg and O seems to 
contradict to the fact that the narrow absorption is absent 
in Mg\,II 7889 \AA, while it is present in O\,I 7773 \AA\ 
 (cf. Anupama et al. 2008).

\subsection{Ejecta, presupernova and explosion}

Our modelling of light curve and other properties of SN~2006jc 
does not permit us to constrain the ejecta mass and explosion energy, 
although models with the moderate energy $(1-2)\times10^{51}$ erg 
are preferred, if one admits 
that the dust forms in the CDS with the velocity of $\sim10^4$ km s$^{-1}$. 
Moreover, models of SN~2006jc with and without $^{56}$Ni 
are plausible, although both have its own drawbacks. The model of a
standard SN~Ibc powered by $^{56}$Ni decay faces 
problems if we identify the blue continuum with the emission of 
the CDS. Only the low mass ejecta $\leq1~M_{\odot}$ could 
reconcile the radioactive model with the proposed scenario 
of the blue continuum.
On the other hand the drawback of the model without $^{56}$Ni 
is that it predicts unacceptably strong X-ray luminosity. 

A more certain judgement on the 
source of SNe~Ibn luminosity at the nebular epoch could be made in principle
on the basis of 
the late time ($t>200$ d) photometry and spectra of SN~Ibn.
A deviation of the bolometric light curve from that predicted 
by the radioactive model would favour the CS interaction as a primary 
energy source. 
On the other hand uniformity of light curves and their correspondance to 
that expected for the radioactive model would favour the radioactivity.
If the radioactivity as a primary energy source will be confirmed the 
second crucial question would remain, whether the ejecta is a standard 
SN~Ibc or the object similar to our low mass model A3 with $\sim1~M_{\odot}$ 
ejecta and $^{56}$Ni mass of $\sim0.3~M_{\odot}$. This issue could be 
elucidated by the late time spectra that may or may not reveal 
strong [O\,I] 6300, 6364 \AA\ emission, the canonical
feature of SN~Ibc, at the nebular stage. The strength of this emission 
relative to Fe\,II line continuum would confirm or reject a picture of 
explosion of a standard SN~Ibc slighly spoiled by CS intraction.
The dust absorption could hamper this test, although it may well be 
that due to clumpiness of the dusty shell the attenuation will not 
be large at the late nebular epoch.

The CSM mass around SN~2006jc at the distance $r\leq2\times10^{15}$ cm 
in our models (Table 1) is in the range of $0.02-0.05~M_{\odot}$.
With the outflow velocity $v_w=600$ km s$^{-1}$ this mass had to be lost 
by pre-SN during the last year before the explosion 
with the rate of $\sim0.02-0.05~M_{\odot}$ yr$^{-1}$ and 
kinetic luminosity of $\sim(2.5-6)\times10^{39}$ erg s$^{-1}$. 
The outflow characteristics suggest extraordinary behavior of pre-SN. The 
pre-SN mass loss might be related with 
the bright outburst two years before the SN~2006jc explosion 
(Nakano et al. 2006; Foley et al. 2007; Pastorello et al. 2007).
However, if the ejection episode was as brief as the optical
outburst then the CS shell should be narrow and reside 
at the distance of about $4\times10^{15}$ cm in contrast to our 
model which suggests a significant fraction of CSM 
in the close vicinity, at $r<2\times10^{15}$ cm. Assuming the 
CS shell expansion velocity of 2400 km s$^{-1}$  one gets the 
CS shell even farther out, at the distance of $\sim1.6\times10^{16}$ cm
(Mattila et al. 2008).

A problem of a vigorous mass loss that preceeds 
 the SN explosion is familiar in relation with SNe~IIn.
Following the conjecture invoked for SN~1994W (Chugai et al. 2004b) one 
might suggest that in some cases, including SNe~Ibn, the mass ejection 
by pre-SN is initiated by the flash of nuclear burning, e.g., of neon. 
However, the current theory of stellar evolution 
does not predict the violent mass loss of pre-SNe 
with initial masses $<100~M_{\odot}$ (Heger et al. 2003). 
Only models of massive stars in the range of $100-140~M_{\odot}$ reveal 
the instability that could result in the strong nuclear outbursts 
 and mass ejection (Heger et al. 2003). The physics is 
 in pair production, which leads to infall, explosion of 
 carbon and oxygen followed by mass ejection. These stars
can experience  several ejection episodes before the ultimate collapse and  
 collisions between consecutively 
 ejected shells could produce optical outburst resemblant supernovae 
events (Heger et al. 2003). 
 Recently, extremely luminous SN~2006gy was identified  with a similar
  phenomenon  (Woosley et al. 2007). 
It is not clear, whether such a scenario is applicable to 
 SNe~Ibn. If does, then these supernovae should be 
 attributed to the CS interaction in the absence of $^{56}$Ni.
 An important implication of this scenario is that the luminous massive 
 helium star should remain at the position of SN~Ibn.

An alternative scenario, in which SN explosion might be preceeded by 
the vigorous mass loss, is a binary system in which 
a black hole experiences a merging with the helium core of 
a red supergiant. 
The supernova explosion in this case is presumably driven by the 
non-relativistic jets generated by the rapid disk accretion of the 
helium core material onto the black hole. In fact this possibility 
is a version of a merger scenario of gamma-ray bursts 
(Fryer \& Woosley 1998; Fryer et al. 1999).
The merging of a neutron star with a red supergiant leads to the 
similar scenario because the inspiraling neutron star that accrete 
at the Bondi-Hoyle accretion rate grows black hole already in the 
hydrogen envelope (Chevalier 1993; Fryer et al. 1999).
The binary scenario predicts asphericity of ejecta caused by jets 
and of the CSM caused by the equatorial outflow from the common envelope.

\section{Conclusions}

The goal of this study was to present a model that might
account for the unusual properties of SNeIbn including SN~2006jc. 
I argue that 
the CS interaction with a dense CSM is essential energy source for the 
early light curve of SN~Ibn. I demonstrated that the 
 CS interaction at the early stage leads to the formation of the CDS 
 which in turn is responsible for the smooth continuum of 
early SN~2000er and unusual blue bumpy continuum of SN~2006jc. 
I showed that the dust can 
form in the CDS at about day 50 in accord with observations.
The modelling of the He\,I line profile affected by the dust absorption 
confirms a picture in which the line-emitting gas is associated with 
the CS clouds shocked and fragmented in the forward shock, while the 
dust is related with the fragmented CDS.  

It should be stressed that scenario in which SN~Ibn supernova is the 
explosion of a standard or luminous SN~Ibc sligtly modified by the 
CS interaction remains unconfirmed and 
photometry and sectroscopy of SNe~Ibn at the late nebular stage 
are needed to clarify the issue. 

\section*{Acknowledgments} 
I am grateful to Ryan Foley for providing me spectra of SN~2006jc.


\end{document}